\documentclass[english,manuscript]{aastex}
\usepackage[T1]{fontenc}
\usepackage[latin9]{inputenc}
\usepackage{array}
\usepackage{float}
\usepackage{bm}
\usepackage{amstext}
\usepackage{graphicx}
\usepackage{esint}

\makeatletter

\providecommand{\tabularnewline}{\\}

\usepackage{aas_macros}

\makeatother

\usepackage{babel}
\begin{document}

\title{Finite-frequency sensitivity kernels in spherical geometry for time-distance
helioseismology }

\author{Krishnendu Mandal$^{1}$, Jishnu Bhattacharya$^{1}$, Samrat Halder$^{2}$
\& Shravan M. Hanasoge$^{1}$}

\affil{$^{1}$Tata Institute of Fundamental Research, Mumbai, India\\
$^{2}$Indian Institute of Technology, Kharagpur, India}
\begin{abstract}
The inference of internal properties of the Sun from surface measurements
of wave travel times is the goal of time-distance helioseismology.
A critical step toward the accurate interpretation of travel-time
shifts is the computation of sensitivity functions linking seismic
measurements to internal structure. Here we calculate finite-frequency
sensitivity kernels in spherical geometry for two-point travel-time
measurements. We numerically build Green's function by solving for
it at each frequency and spherical-harmonic degree and summing over
all these pieces. These computations are performed in parallel (``embarrassingly''),
thereby achieving significant speedup in wall-clock time. Kernels
are calculated by invoking the first-order Born approximation connecting
deviations in the wavefield to perturbations in the operator. Validated
flow kernels are shown to produce travel-times within $0.47\%$ of
the true value for uniform flows up to $750\,\text{m}/\text{s}$.
We find that travel-time can be obtained with errors of $1$ millisecond
or less for flows having magnitudes similar to meridional circulation.
Alongside flows, we also compute and validate sensitivity kernel for
sound-speed perturbations. These accurate sensitivity kernels might
improve the current inferences of sub-surface flows significantly. 
\end{abstract}

\section{Introduction\label{introduction}}

Seismic waves are observed on the solar surface by studying Doppler
shifts of specific spectral lines produced in the photosphere. These
waves are produced by vigorous turbulence near the solar surface,
and they travel through the solar interior before resurfacing. Measuring
the wave velocity field on the surface therefore opens up a window
into the solar subsurface that is otherwise opaque to electromagnetic
observations. Seismic waves are sensitive to subsurface features that
either change the wave speed; in turn, information gleaned from surface
observations of these waves can be inverted to image the interior
that the wave has traversed. Local helioseismology can be used to
infer, among other things, flows of various length scales inside the
Sun, magnetic fields and active regions, and thermal anomalies leading
to deviations in sound speed from that in the stratified hydrostatic
background. 

There are various approaches of relating seismic observations to subsurface
features (for reviews see e.g. \citet{2005LRSP....2....6G,2010ARA&A..48..289G,2016AnRFM..48..191H}),
one among them being time-distance helioseismology \citep{duvall1993}
where we relate travel-time maps on the solar surface --- obtained
from wave cross-correlations --- to interior features . Wave travel-times,
as measured on the surface, will change if the wave encounters sound-speed
perturbations or flows as it passes through the solar interior. Among
several other approaches to relate change in travel times with perturbations
in the background medium, the formalism proposed by \citet{2000SoPh..192..193B,2002ApJ...571..966G}
using first-order Born approximations has been widely adopted. Key
to the relationship between travel time shifts and perturbations in
the medium is travel-time sensitivity kernel which describes how sensitive
travel times are to changes in model parameters. Several authors,
e.g. \citet{2007ApJ...671.1051J,2007AN....328..228B,2015SSRv..196..201B}
have used the formalism of \citet{2002ApJ...571..966G} to compute
sensitivity kernels for sound speed and flows in Cartesian geometry.
Cartesian formulations of the inverse problem are limited to spatial
scales much smaller than the solar radius e.g. for the studies of
sunspots, supergranulation etc. 

Since the Sun is spherical, it is important to extend this formalism
to spherical geometry to reliably image large-scale structures e.g.
meridional flows \citep{1979SoPh...63....3D,1997Natur.390...52G},
differential rotation, tachocline etc. Computing these kernels is
expensive and due to this limitation, several authors e.g. \citet{2013ApJ...774L..29Z,2015ApJ...805..133J,2015ApJ...813..114R}
have applied the ray approximation in place of the first-order Born
approximation to compute flow-sensitivity kernels. Ray theory is an
infinite frequency limit in which the travel time is sensitive only
to perturbations along the ray path. Results from ray theory are reliable
only if the length scale of the perturbation is significantly greater
than the wavelength \citep{2001ApJ...561L.229B,2004ApJ...616.1261B}.
Since the length scales over which perturbations vary are not known
a priori in these inverse problems, it is important to perform inversions
using the best-possible kernels. Recently, \citet{2016ApJ...824...49B,2016arXiv161101666G}
have computed sensitivity kernels in spherical geometry. \citet{2016ApJ...824...49B}
use a normal-mode expansion to compute Green's function. This approach
converges slowly and is therefore computationally expensive (personal
communication, A. C. Birch, \citet{2016arXiv161101666G}). \citet{2016arXiv161101666G}
reduce a gravity-free wave equation to a scalar equation and solve
it using a finite element analysis method in an axisymmetric background. 

In this work, we propose a different approach. We follow the measurement
process described in \citet{2002ApJ...571..966G} to derive expressions
for sensitivity kernels for sound-speed, flows and stream function
in terms of Green's function and its derivative. We numerically solve
for Green's functions in a spherically symmetric background using
a finite-difference based scheme and compute kernels with high accuracy.
We also show kernels for an azimuthal stream function which takes
into account continuity and therefore appropriate for meridional-flow
inversions. Since kernels are computed about a spherically symmetric
background, so the inversions have to be linear, but we show that
linearity is a good assumption for flows having magnitudes similar
to meridional circulations.

\section{Computing Green's function\label{green}}

We consider a temporally stationary, spherically symmetric, non-rotating,
non-magnetic solar model at hydrostatic equilibrium parametrized through
material composition and thermodynamic properties at each point. Assuming
spherical symmetry, material properties such as density and acceleration
due to gravity, and thermal properties such as pressure and sound-speed
depend only on the radial distance $r$ from the center of the Sun.
We choose Model S \citep{ChristensenDalsgaard:1996ap} as our background
solar model. In further analysis, we use the symbol $\rho_{0}\left(r\right)$
to denote the radial density profile, $p_{0}\left(r\right)$ to denote
the radial pressure profile, $\mathbf{g}_{0}\left(r\right)$ to denote
acceleration due to gravity and $c\left(r\right)$ to denote the sound-speed.
Seismic waves result in small deviations of these parameters about
their equilibrium values, we denote these deviations using unsubscripted
and primed variables. $\bm{\xi}(\mathbf{r},\omega)$ which is displacement
vector of seismic waves follows the wave equation, where $\omega$
is temporal frequency, 
\begin{equation}
-\rho_{0}(\mathbf{r})\left(\omega+i\gamma\right)^{2}\bm{\xi}(\mathbf{r},\omega)=-\bm{\nabla}p^{\prime}(\mathbf{r},\omega)+\rho^{\prime}(\mathbf{r},\omega)\mathbf{g}_{0}(\mathbf{r})+\mathbf{F}(\mathbf{r},\omega),\label{eq:wave_eqn}
\end{equation}
where $\mathbf{F}\left(\mathbf{r},\omega\right)$ denotes sources
excitation, $\gamma$ is attenuation. $p^{\prime}$ and $\rho^{\prime}$
pressure and density perturbation respectively. Splitting (\ref{eq:wave_eqn})
into tangential and radial components, we obtain 
\begin{eqnarray}
\partial_{r}p^{\prime} & = & \rho_{0}(\omega+i\gamma)^{2}\xi_{r}-c^{-2}p^{\prime}\mathbf{g}_{0}-\rho_{0}\xi_{r}N^{2}g_{0}+F_{r},\label{eq:radial_part}\\
\bm{\nabla}_{h}p^{\prime} & = & \rho_{0}\omega^{2}\bm{\xi}_{h}+\mathbf{F}_{h},\label{eq:tangential_part}
\end{eqnarray}
where $\bm{\nabla}_{h}$ represents the lateral component of the gradient
$\bm{\nabla}$, and $N$ is the Brunt-V\"ais\"al\"a frequency.
The perturbed parameters are also constrained by the continuity equation,
\begin{equation}
\rho^{\prime}=-\frac{1}{r^{2}}\partial_{r}\left(r^{2}\rho_{0}\xi_{r}\right)+\rho_{0}\mbox{\ensuremath{\bm{\nabla}}}_{h}\cdot\bm{\xi}_{h}.\label{eq:continuity}
\end{equation}
We also assume that the perturbations are adiabatic in nature, so
the pressure perturbation $p^{\prime}$ and the density perturbation
$\rho^{\prime}$ and radial displacement $\xi_{r}$ are related through
\begin{equation}
\rho^{\prime}=\frac{p^{\prime}}{c^{2}}+\frac{\rho_{0}}{g_{0}}N^{2}\xi_{r}.
\end{equation}
This set of equations forms a well-determined system that we solve
for quantities $\bm{\xi}$, $\rho^{\prime}$ and $p^{\prime}$. We
simplify the system by eliminating $\rho^{\prime}$ and the tangential
components of $\bm{\xi}$, therefore reducing the system to two equations
in two unknowns: $\xi_{r}$ and $p^{\prime}$. 

The displacement vector, $\bm{\xi}(\mathbf{r},\omega)$ is related
to Green's function through 
\begin{equation}
\xi_{i}(\mathbf{r},\omega)=\int G_{ij}(\mathbf{r},\mathbf{r}^{\prime},\omega)F_{j}(\mathbf{r}^{\prime},\omega)d\mathbf{r}^{\prime},\label{eq:green_xi}
\end{equation}
where indices $i$, $j$ denote $r,\,\theta,\,\phi$. We use Einstein's
summation convention here. $G_{ij}(\mathbf{r},\mathbf{r}^{\prime})$
is the seismic response of the $j$ th component of the point source,
located at $\mathbf{r}^{\prime}$, measured in the $i$ th component
of the displacement vector, at position $\mathbf{r}$. In order to
obtain Green's function, a radially directed point source, placed
at $\mathbf{r}_{s}$ is considered as a source function in the wave
equation 
\begin{equation}
F_{i}(\mathbf{r},\omega)=\delta(\mathbf{r}-\mathbf{r}_{s})\delta_{ir}.\label{eq:source_func}
\end{equation}
 Applying Equation (\ref{eq:source_func}) to Equation (\ref{eq:green_xi}),
we obtain 
\begin{eqnarray}
\xi_{r}(\mathbf{r},\omega) & = & G_{rr}(\mathbf{r},\mathbf{r}_{s},\omega),\qquad\xi_{\theta}(\mathbf{r},\omega)=G_{\theta r}(\mathbf{r},\mathbf{r}_{s},\omega),\qquad\xi_{\phi}(\mathbf{r},\omega)=G_{\phi r}(\mathbf{r},\mathbf{r}_{s},\omega),\label{eq:green_1}
\end{eqnarray}
which means that the radial and horizontal components of the displacement
vector $\bm{\xi}$ for a radially directed delta function point source
describe Green's function $G_{rr}$ and $\mathbf{G}_{hr}$ respectively,
where $\mathbf{G}_{hr}=\left(G_{\theta r},G_{\phi r}\right)$. We
expand $\xi_{r}$, $p^{\prime}$ and source $F_{r}$ in the spherical-harmonic
basis 
\begin{eqnarray}
\xi_{r}(\mathbf{r};\omega) & = & \sum_{\ell m}\alpha_{\ell\omega}(r)Y_{\ell m}(\theta,\phi)Y_{\ell m}^{*}(\theta_{s},\phi_{s}),\nonumber \\
p^{\prime}(\mathbf{r};\omega) & = & \sum_{\ell m}\beta_{\ell\omega}(r)Y_{\ell m}(\theta,\phi)Y_{\ell m}^{*}(\theta_{s},\phi_{s}),\nonumber \\
F_{r}(\mathbf{r},\omega) & = & \sum_{\ell m}\delta(r-r_{s})Y_{\ell m}(\theta,\phi)Y_{\ell m}^{*}(\theta_{s},\phi_{s}),\label{eq:spherical_harmonics}
\end{eqnarray}
where $Y_{\ell m}(\theta,\phi)$ is the spherical harmonic of degree
$\ell$ and azimuthal order $m$. Substituting Equation (\ref{eq:spherical_harmonics})
into Equation (\ref{eq:radial_part}) and (\ref{eq:tangential_part}),
we obtain a coupled system of ordinary differential equations
\begin{equation}
\mathrm{M}\left(\begin{array}{c}
\alpha_{\ell\omega}\left(r\right)\\
\beta_{\ell\omega}\left(r\right)
\end{array}\right)=\left(\begin{array}{c}
0\\
\delta(r-r_{s})
\end{array}\right),\label{eq:matrix_eqn}
\end{equation}
where 
\begin{eqnarray}
\mathrm{M} & = & \left(\begin{array}{cc}
\frac{d}{dr}-(\frac{g_{0}}{c^{2}}-\frac{2}{r}) & -\frac{1}{\rho_{0}c^{2}}\left(\frac{\ell(\ell+1)c^{2}}{r^{2}}-1\right)\\
\rho_{0}((\omega+i\gamma)^{2}-N^{2}) & \frac{d}{dr}+\frac{g_{0}}{c^{2}}
\end{array}\right).
\end{eqnarray}
Equation (\ref{eq:matrix_eqn}) has to be solved numerically as a
function of radius for each temporal frequency $\omega$ and harmonic
degree $\ell$, yielding the pair $\left(\alpha_{\ell\omega}\left(r\right),\beta_{\ell\omega}\left(r\right)\right)$.
Using Equation (\ref{eq:green_1}), we construct components of Green's
function from $\alpha_{\ell\omega}$ and $\beta_{\ell\omega}$, 
\begin{eqnarray}
G_{rr}(\mathbf{r},\mathbf{r}_{s},\omega) & = & \sum_{\ell}\frac{(2\ell+1)}{4\pi}\alpha_{\ell\omega}(r)P_{\ell}(\cos(\hat{\mathbf{r}}\cdot\hat{\mathbf{r}}_{s})),\label{eq:green_expn2}\\
\mathbf{G}_{hr}(\mathbf{r},\mathbf{r}_{s},\omega) & = & \sum_{\ell}\frac{(2\ell+1)}{4\pi\omega^{2}\rho_{0}}\beta_{\ell\omega}(r)\bm{\nabla}_{h}P_{\ell}(\cos(\hat{\mathbf{r}}\cdot\hat{\mathbf{r}}_{s})).\label{eq:green_expn}
\end{eqnarray}

\subsection{Model for wave damping}

\begin{figure}[H]
\centering{}\includegraphics[scale=0.5]{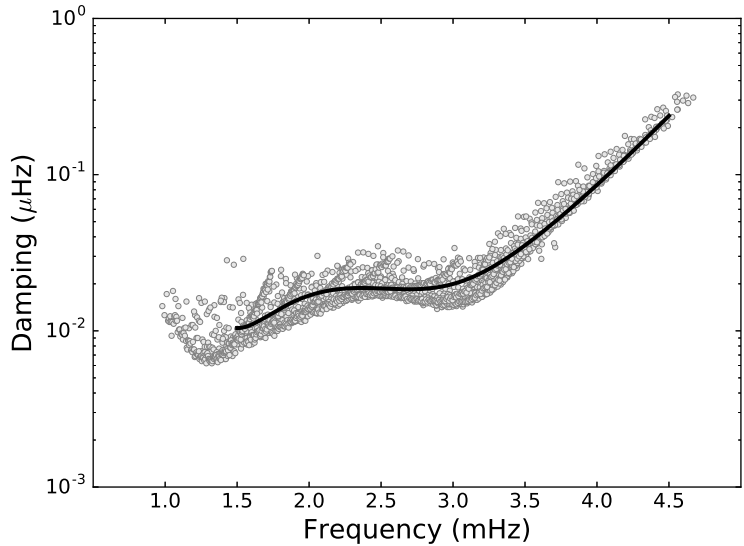}\caption{\label{fig:damping}Line-widths for modes with harmonic degree $\ell$
lying between $11$ and $200$ (grey markers) \citep{Schou1999}.
The line widths correspond to damping timescales and depend primarily
on mode frequency. We find that this frequency dependence can be approximated
by a sixth-order polynomial. The best fit polynomial for $\ell=30$
has been plotted in black. We use this functional form of the damping
scale in our analysis.}
\end{figure}

Waves in the Sun have finite lifetimes, and are attenuated over a
period of a few days. The decay of modes results from dynamical origins
such as coupling with turbulent pressure and leakage into the atmosphere,
as well as thermal ones such as radiative losses and interaction of
waves with turbulent heat flux \citep[see][ and references therein]{Houdek99,2015ApJ...806..246B}.
Damping of wave modes is usually modeled by adding a small imaginary
component to the mode frequency, that is by setting $\omega_{n\ell}=\omega_{n\ell}^{0}+i\gamma$,
where $\omega_{n\ell}^{0}$ represents the frequency of the ideal
adiabatic undamped wave. Observational studies \citep{Schou1999}
show that the damping parameter $\gamma$ is primarily dependent on
mode eigenfrequency $\omega_{n\ell}$, and to a lesser extent on the
harmonic degree $\ell$ of the mode. We have plotted the measured
damping parameter as a function of frequency in Fig. \ref{fig:damping}.
Ignoring the $\ell$-dependence of $\gamma$, we find that it can
be approximately represented as a sixth-order polynomial of frequency
as 
\begin{equation}
\gamma=a_{0}+a_{1}\omega+a_{2}\omega^{2}+a_{3}\omega^{3}+a_{4}\omega^{4}+a_{5}\omega^{5}+a_{6}\omega^{6},\label{eq:damping_power}
\end{equation}
where the value of the coefficients are noted in Table (\ref{tab:damping}). 

\begin{table*}
\label{table1}%
\begin{tabular}{|>{\centering}p{2cm}|>{\centering}p{2cm}|>{\centering}p{2cm}|>{\centering}p{2cm}|>{\centering}p{2cm}|>{\centering}p{2cm}|>{\centering}p{2cm}|}
\hline 
$a_{0}$

$(\mu\text{Hz})$ & $a_{1}$

$(\mu\text{Hz})^{0}$ & $a_{2}$

$(\mu\text{Hz})^{-1}$ & $a_{3}$

$(\mu\text{Hz})^{-2}$ & $a_{4}$

$(\mu\text{Hz})^{-3}$ & $a_{5}$

$(\mu\text{Hz})^{-4}$ & $a_{6}$

$(\mu\text{Hz})^{-5}$\tabularnewline
\hline 
$1.33\times10^{-6}$ & $-3.20\times10^{-3}$ & $3.12$ & $-1.57\times10^{3}$ & $4.30\times10^{5}$ & $-6.19\times10^{7}$ & $3.69\times10^{9}$\tabularnewline
\hline 
\end{tabular}

\caption{\label{tab:damping}Values of the coefficients in the polynomial expansion
of damping scale $\gamma$ (Equation (\ref{eq:damping_power})).}
\end{table*}

\subsection{Boundary conditions\label{boundary}}

The system in Equation (\ref{eq:matrix_eqn}) has to be augmented
with appropriate boundary condition to obtain solutions. We are interested
in trapped modes, that is waves with frequency lying in the range
$2\,\text{mHz}$ to $5.5\,\text{mHz}$; these waves are reflected
back into the solar interior at the surface. The inwards reflection
takes place because of a sharp increase in the acoustic cutoff frequency
close to the surface. While propagating into the interior, seismic
waves are refracted away from the center because of increasing sound
speed, and at a specific depth --- referred to as the turning point
--- these modes are totally internally reflected back towards the
solar surface. The depth at which total internal reflection occurs,
depends on the frequency and horizontal wavenumber \textbf{$k_{\text{h}}=\sqrt{\ell(\ell+1)}/R_{\odot}$}.
This picture of waves being totally reflected back, however, is inherently
ray-theoretic in nature; waves of a finite frequency are exponentially
damped beyond the turning point and have a finite non-zero --- albeit
decaying --- amplitude deeper in the interior. 

We choose $0.2R_{\odot}$ as the inner boundary and we do not consider
modes whose turning points are below $0.2R_{\odot}$. With no loss
of generality, we can push the lower boundary closer to the core.
We assume that waves corresponding to harmonic degrees greater than
$20$ have turning points above $0.2R_{\odot}$ and choose $20$ as
the lower cutoff of harmonic degrees in our analysis. We set the radial
component of wave displacement to zero at the lower boundary, that
is

\begin{equation}
\xi_{r}(r=0.2R_{\odot},\theta,\phi;\omega)=0.\label{eq:boundary_1}
\end{equation}

Beyond the outer surface, the waves with frequencies below the acoustic
cutoff are exponentially damped. The pressure perturbation corresponding
to the wave would rapidly decay to zero with height, which is why
we peg its value to zero at the upper boundary of our domain, that
is at $r=r_{\text{out}}$ the pressure perturbation $p^{\prime}$
satisfies 
\begin{equation}
p^{\prime}(r=r_{\text{out}},\theta,\phi;\omega)=0.\label{eq:boundary_2}
\end{equation}
 The Equations (\ref{eq:boundary_1}) and (\ref{eq:boundary_2}) hold
for all $(\theta,\,\phi)$ and from Equations (\ref{eq:green_expn})
and (\ref{eq:green_expn2}) that is possible only if
\begin{eqnarray}
\alpha_{\ell\omega}(r=r_{\text{in}}) & = & 0,\nonumber \\
\beta_{\ell\omega}(r=r_{\text{out}}) & = & 0.\label{eq:boundary_11}
\end{eqnarray}
We use boundary condition (\ref{eq:boundary_11}) to solve Equation
(\ref{eq:matrix_eqn}) for $\alpha_{l\omega}$ and $\beta_{l\omega}$.

\subsection{Numerical technique\label{numerical}}

Evaluating Green's function requires us to solve Equation (\ref{eq:matrix_eqn})
for each (discretized) frequency $\omega$ and harmonic degree $\ell$
that encompass the spectrum of solar seismic eigenmodes. We choose
a frequency range from $2\,\text{mHz}$ to $4.5\,\text{mHz}$, split
into $1250$ bins. We choose harmonic degree $\ell$ lying in a range
$\left[20,\ell_{\text{max}}\right]$. The choice of the upper cutoff
$\ell_{\text{max}}$ is primarily governed by the convergence of the
final sensitivity kernel, since increasing the cutoff $\ell_{\text{max}}$
would also necessitate increasing the resolution of the discretized
angular $\left(\theta,\phi\right)$ grid to avoid aliasing while computing
wave travel-times using the kernel. Evaluating Green's function using
Equation (\ref{eq:matrix_eqn}) involves discretizing the radius $r$
and generating the matrix on the left-hand side; each $\left(\omega-\ell\right)$
pair leads to one matrix, leading to one set of solutions $\left(\alpha_{\ell}\left(r,\omega\right),\beta_{\ell}\left(r,\omega\right)\right)$.
We use Model S to evaluate matrix elements. We choose $1596$ radial
points distributed evenly in acoustic distance, leading to matrices
of size $3192\times3192$. Spherical symmetry and linearity dictates
that the solutions for different $\left(\omega,\ell\right)$ pairs
are independent, a fact that we utilize to compute the different solutions
in parallel on a computer cluster. We solve Equation (\ref{eq:matrix_eqn})
using the \textit{linalg} module implemented in \textit{numpy}, and
subsequently evaluate various components of Green's function listed
in Equation (\ref{eq:green_expn}) in $\omega-\ell$ space. We construct
the matrix in Equation (\ref{eq:matrix_eqn}) by discretizing derivatives
on the radial grid using various stencils.

We apply a second-order backward finite difference scheme to evaluate
the first derivative in Equation (\ref{eq:matrix_eqn}) at the boundary
points. Close to the boundary except for boundary points, we use second-order
central differences. Farther away from the boundary, we increase the
accuracy of the central-difference scheme up to sixth order. We approximate
the delta function by the following Gaussian:
\begin{equation}
\delta(r-r_{s})\approx\frac{\text{exp}[-(r-r_{s})^{2}/(2\Delta^{2})]}{\sqrt{2\pi}r^{2}\Delta},
\end{equation}
where $\Delta$ is the width of the function. We have chosen $\Delta=8\,\mathrm{km}$
and we place our source at $75$ km below the surface. The reason
for this particular choice of $\Delta$ is to use $30$ points to
resolve the Gaussian. We have considered $\ell_{\text{max}}=300$
for all the plots of sensitivity kernels in the following sections.

\section{Validation of Green's function\label{validation}}

\subsection{Time-distance diagram}

The primary observation in seismology is the line-of-sight projected
velocity at each point on the solar disk. Waves in the Sun are stochastically
excited by turbulent convection near the surface, and the sources
that excite waves are distributed randomly over the solar disk. In
our analysis, we place a point source and study waves emanating from
it. We record the waves as they pass through specific points on the
surface that we label as ``receivers''. Each source-receiver pair
yields information about the sub-surface medium that the wave travels
through. Waves recorded at each receiver over the entire period of
observation is referred to as a time-distance diagram (for a description
of time-distance diagrams and how they are obtained from observations
of solar disk, see \citet{duvall1993}).

\begin{figure}[H]
\begin{centering}
\includegraphics[scale=0.6]{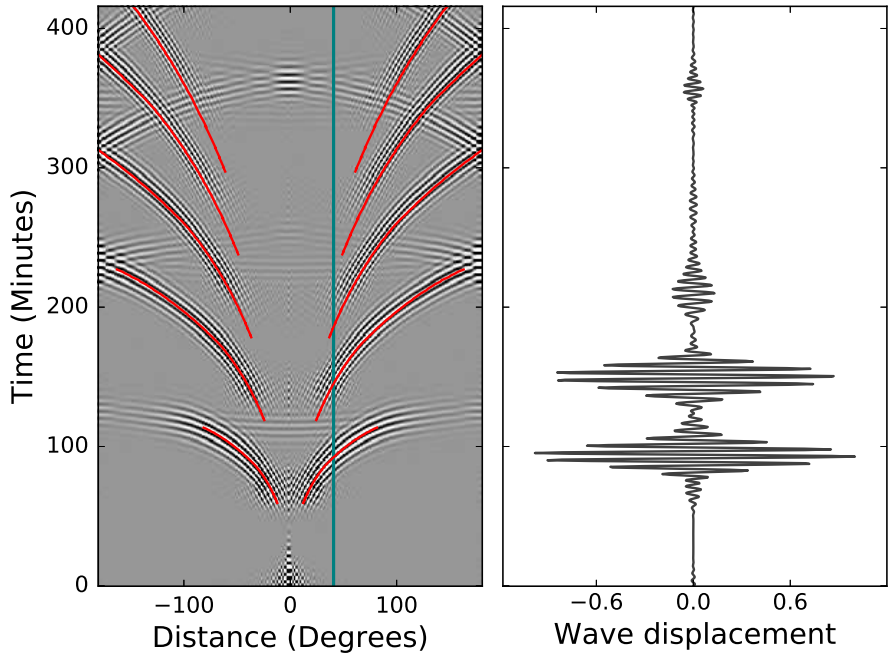}
\par\end{centering}

\caption{\label{fig:time_distance}Left panel: Time-distance diagram computed
from Equation (\ref{eq:source_freq}). Red solid lines are from ray-theory,
computed at a frequency of 3.2 mHz. Right panel: Cut through time-distance
diagram at a receiver position highlighted by a solid vertical line
in the left panel. This plot indicates the arrival of the waves at
the receiver location after encountering different number of bounces
in the solar interior. }
\end{figure}

The time-distance diagram acts as a validation test for Green's function
as we may compare it with the diagram obtained separately in the ray
theory limit. In our case, we study the wave displacement instead
of velocity, the former being a time-integral of the latter. The wave
displacement is given by

\begin{equation}
\xi_{i}(\mathbf{r},t)=\int_{-\infty}^{\infty}dt\,G_{ir}(\mathbf{r},\mathbf{r}_{s},\omega)F_{r}(\mathbf{r}_{s},\omega)e^{i\omega t}.\label{eq:source_freq}
\end{equation}
We assume a Gaussian frequency dependence of the source, that is
\begin{equation}
F(\mathbf{r}_{s},\omega)=\text{\ensuremath{\exp}}\left(-\frac{(\omega-\omega_{0})^{2}}{2\sigma^{2}}\right),\label{eq:source_dist}
\end{equation}
where $\omega_{0}=2\pi\times3.2$ mHz, and $\sigma=2\pi\times0.4$
mHz. We use the same parameters for the computation of kernel. We
compare the time-distance diagram from our simulation with ray-theory
(e.g. \citet{2000PhDT.........9G}) in Fig. \ref{fig:time_distance}. 

\subsection{Power spectrum\label{power_spec}}

We compute the power spectrum of the waveform in temporal and spatial
frequency space. Time series of velocity amplitudes of seismic waves,
recorded by the Michelson Doppler Imager (MDI, \citealp{Scherrer95})
onboard the Solar and Heliospheric Observatory (SOHO, \citealp{Domingo95}),
have been used to generate high-resolution seismic power spectra \citep{Rhodes1997,rhodes98,Schou1999}.
This provides us with a ready test for Green's functions, in that
the resonant ridges in the numerically computed function should match
those observed in the Sun.

\begin{figure}[H]
\begin{centering}
\includegraphics[scale=0.45]{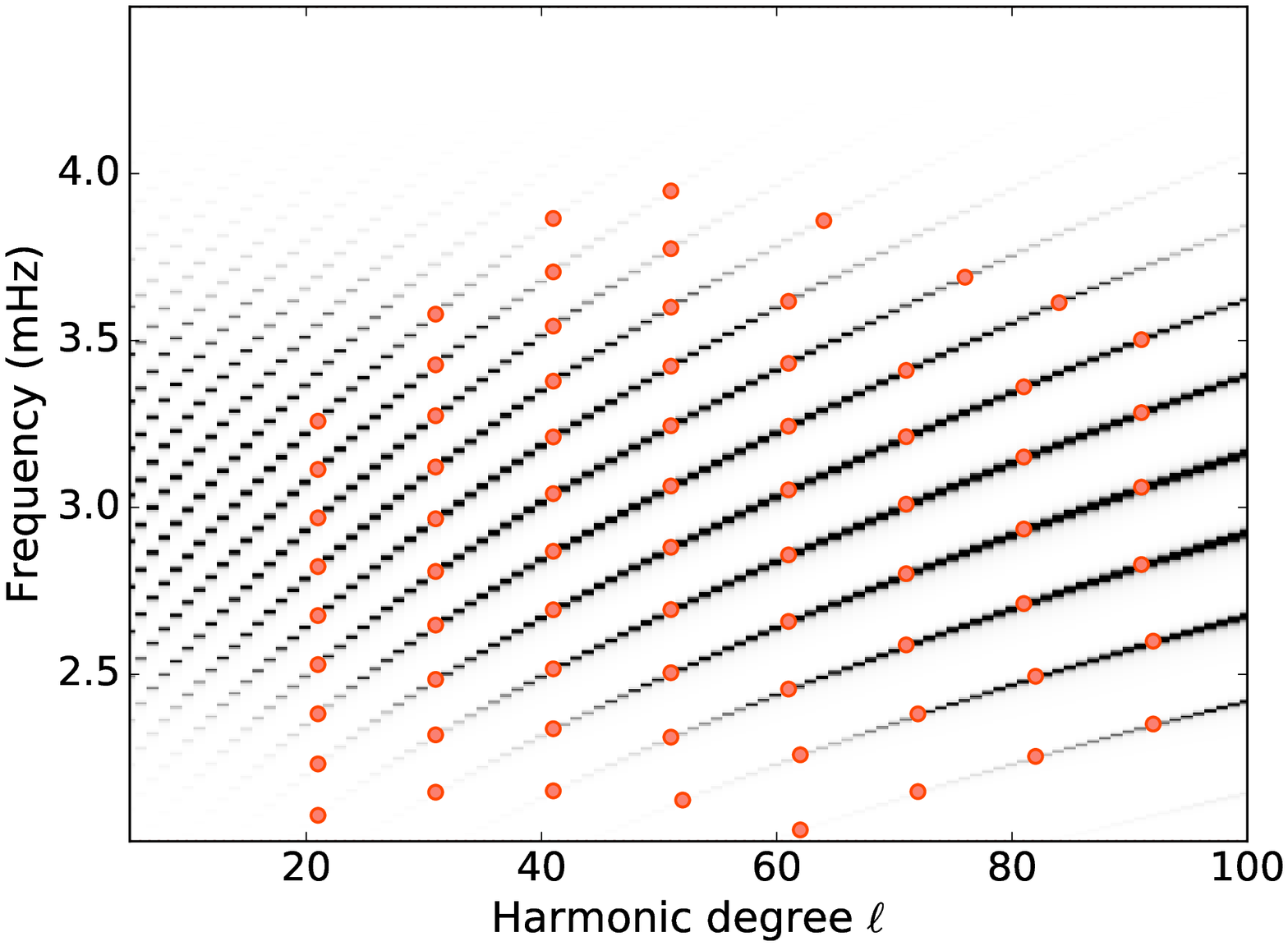} \includegraphics[scale=0.45]{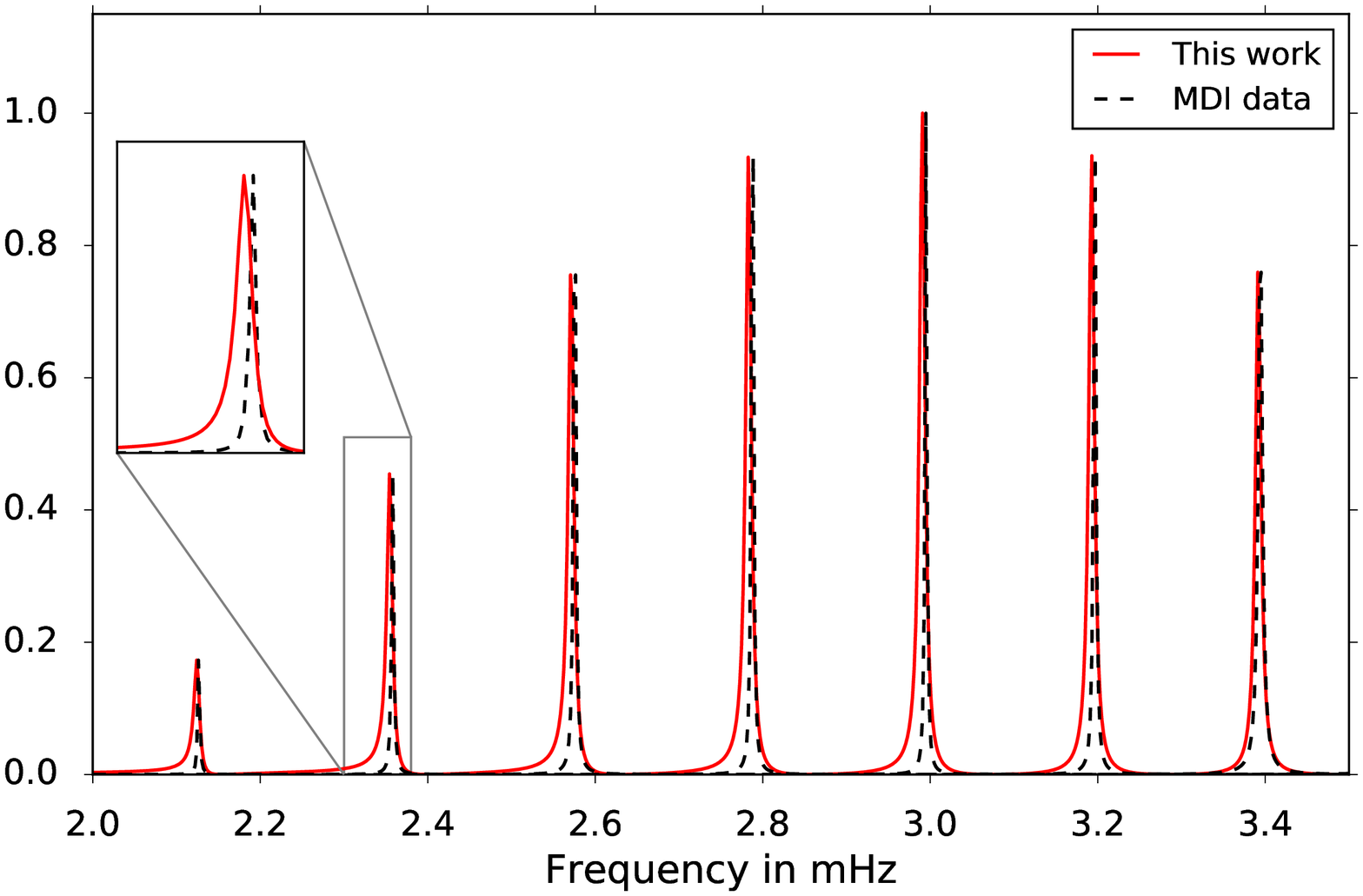}
\par\end{centering}

\caption{\label{fig:pow_spec}Upper panel: Power spectrum computed from Equation
(\ref{eq:power_spectrum}). Red `o' markers are eigenmodes obtained
from 72 days MDI observation by \citet{Schou1999}. Lower Panel: Power
spectrum for harmonic degree $\ell=70$ from our simulation is shown
by red solid line. Black dashed lines are Lorentzians whose centers
and widths are the eigenmodes and corresponding FWHM respectively,
obtained from a 72-day time series of MDI observation by \citet{Schou1999}.
The peaks of the Lorentzian have been normalized to the amplitude
of the nearest peak of the model power spectrum. In the inset, one
of the peaks has been zoomed into. The mismatch between simulated
and measured frequencies may be attributed to inaccuracies in our
choice of surface boundary conditions as compared to the Sun \citep{2001ApJ...561.1127R}
and the imperfect modeling of surface layers in model S \citep{1999A&A...351..689R}.}
\end{figure}

The first step in computing the power spectrum is to carry out a spherical
harmonic transform of the wave displacement to obtain 
\begin{equation}
\xi_{\ell m}\left(r_{\text{obs}};\omega\right)=\int\xi_{r}\left(r_{\text{obs}},\theta,\phi;\omega\right)Y_{\ell m}(\theta,\phi)d\Omega,
\end{equation}
where $d\Omega$ is the spherical solid angle and $r_{\text{obs}}$
is the radial coordinate of the height at which observations are carried
out. For simplicities, we set $r_{\text{obs}}=r_{s}$ here. Since
the background model is spherically symmetric, the spectrum does not
depend on the azimuthal degree $m$, therefore we average over it
to obtain power at each angular mode $\ell$. The $m-$averaged power
spectrum of the wave displacement is given by 
\begin{equation}
P_{\ell}\left(r_{\text{obs}};\omega\right)=\frac{1}{2\ell+1}\sum_{m=-\ell}^{\ell}\left|\xi_{\ell m}\left(r_{\text{obs}};\omega\right)\right|^{2}.\label{eq:power_spectrum}
\end{equation}
In Fig.~\ref{fig:pow_spec}, we compare the numerical spectrum computed
from our analysis with that obtained from $72$ days MDI mode-parameter
measurements by \citet{Schou1999}. We notice small mismatch between
simulated and measured frequencies in Fig.~\ref{fig:pow_spec}. This
may be attributed to inaccuracies in our choice of surface boundary
conditions as compared to the Sun \citep{2001ApJ...561.1127R} and
the imperfect modeling of surface layers in model S \citep{1999A&A...351..689R}.

\section{Sensitivity kernels\label{kernel}}

A change in the background that the wave propagates through results
in a variation in seismic waves measured at the surface. This in turn
may change in the wave travel times between the source and receivers.
\citet{2002ApJ...571..966G} developed a technique to compute travel
times from wave cross-correlations by minimizing the misfit between
the observed and model cross correlations. Their formulation, however,
is not specific to cross-correlations and can be extended to other
wave measurables. The use of cross-correlations is necessary for solar
observations since the measured wave velocity is inherently a stochastic
quantity. This is because the location of sources and excitation of
waves is random. In our analysis, however, we assume that the location
and excitation of the wave source is entirely deterministic. Under
this assumption, we relate the wave displacement directly to travel-time
shifts. Denoting the radial component of wave displacement in spherically
symmetric Model S by $\xi_{r}^{0}$ and that in a different background
--- possibly with reduced symmetry --- by $\xi_{r}$, the difference
in source-receiver travel times for these two wavefields can be expressed
as

\begin{equation}
\delta\tau(\mathbf{r}_{r},\mathbf{r}_{s})=\int_{-\infty}^{\infty}dt\,h(t)(\xi_{r}(\mathbf{r}_{r},\mathbf{r}_{s},t)-\xi_{r}^{0}(\mathbf{r}_{r},\mathbf{r}_{s},t)),\label{eq:travel_time}
\end{equation}
where $\mathbf{r}_{r}$ and $\mathbf{r}_{s}$ are the receiver and
source location respectively, and the function $h(t)$ is defined
as 
\begin{equation}
h(t)=\frac{-W(t)\dot{\xi}_{r,0}(\mathbf{r}_{r},\mathbf{r}_{s},t)}{\int_{-\infty}^{\infty}dt^{\prime}W(t)\left[\dot{\xi}_{r,0}(\mathbf{r}_{r},\mathbf{r}_{s},t)\right]^{2}},
\end{equation}
where $W(t)$ is the window function that, in our case, selects only
the first arrival of the waves at the receiver point $\mathbf{r}_{r}$. 

The background model can change because of various reasons, for example
a local bump in the thermal properties resulting in an altered sound
speed, or there being small or large scale flows that the waves propagate
through and are advected by. These perturbations will leave their
imprint on wave travel times. Key to seismic inference is a linear
relation between wave travel-times and the model perturbation. Given
a generic three-dimensional local perturbation $\delta q\left(\mathbf{r}\right)$
in the solar model, the impact it has on the travel time can be quantified
as 
\begin{equation}
\delta\tau=\int_{\odot}K_{q}(\mathbf{r})\delta q\left(\mathbf{r}\right)\,d\mathbf{r},\label{eq:kernel_travel_time}
\end{equation}
where $K_{q}$ is referred to as the sensitivity kernel. This kernel
encodes information about the local impact of a perturbation on measured
travel times. Viewed from the vantage of an inverse problem, the kernel
also represents the gradient of travel-times in the parameter-space
of the perturbation $\delta q$. In the first-order Born approximation,
the sensitivity kernel $K_{q}$ obeys 
\begin{equation}
K_{q}(\mathbf{r})\delta q(\mathbf{r})=\int d\omega\,G_{rj}(\mathbf{r}_{r},\mathbf{r},\omega)\left[\delta\mathcal{L}G(\mathbf{r},\mathbf{r}_{s},\omega)\right]_{jr}h^{*}(\omega)F(\mathbf{r}_{s},\omega),\label{eq:gen_expn}
\end{equation}
where $\delta\mathcal{L}$ is the change in wave operator $\mathcal{L}$
due to the change in parameter $q$ and $h^{*}(\omega)$ is the complex
conjugate of the Fourier transform of the function $h(t)$. In this
work, we propose an efficient way to evaluate sensitivity kernels
in spherical geometry.

\subsection{Sensitivity kernel for sound speed\label{sub:sound-speed-sec}}

\begin{figure*}
\includegraphics[scale=0.45]{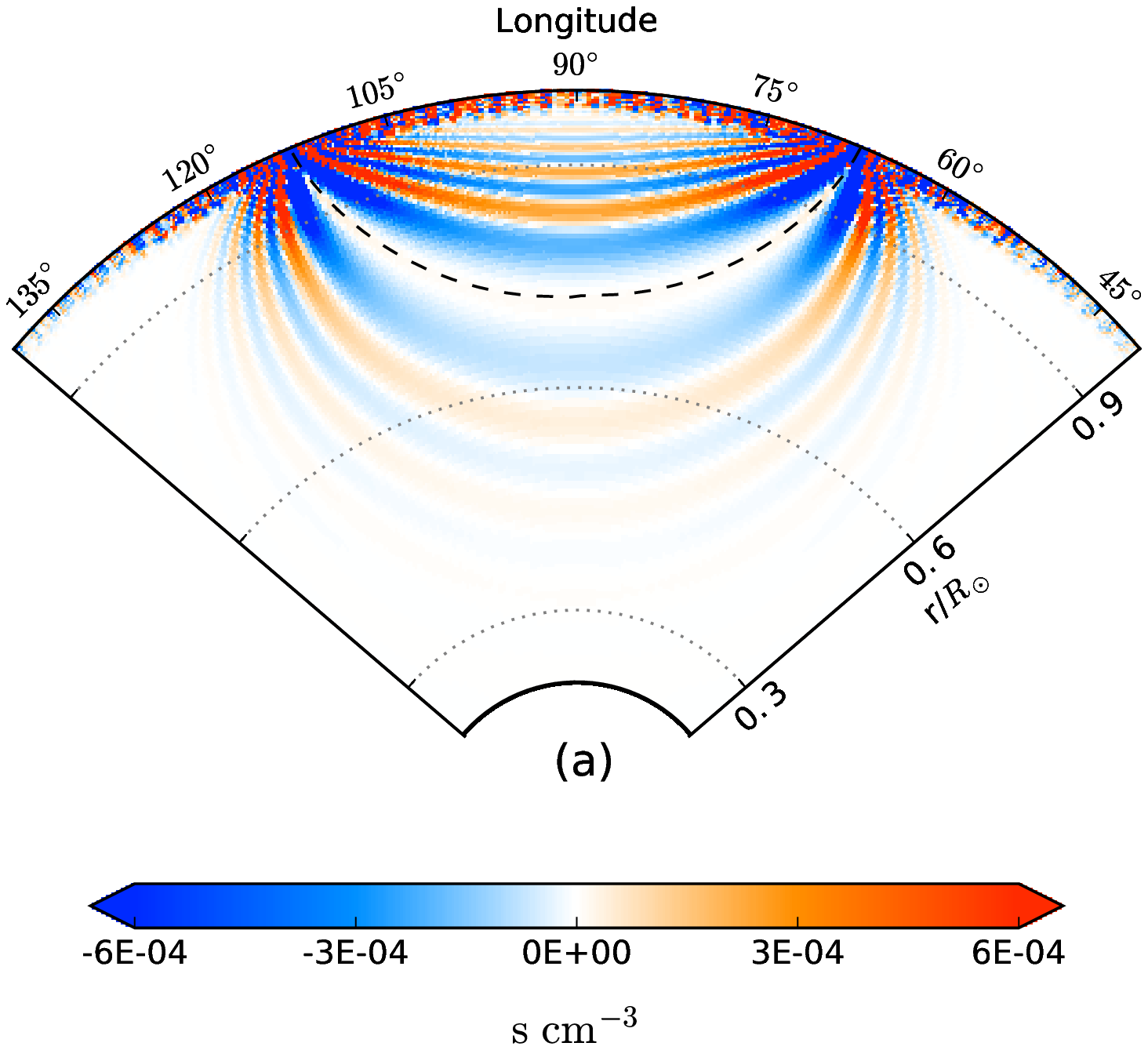}\includegraphics[scale=0.45]{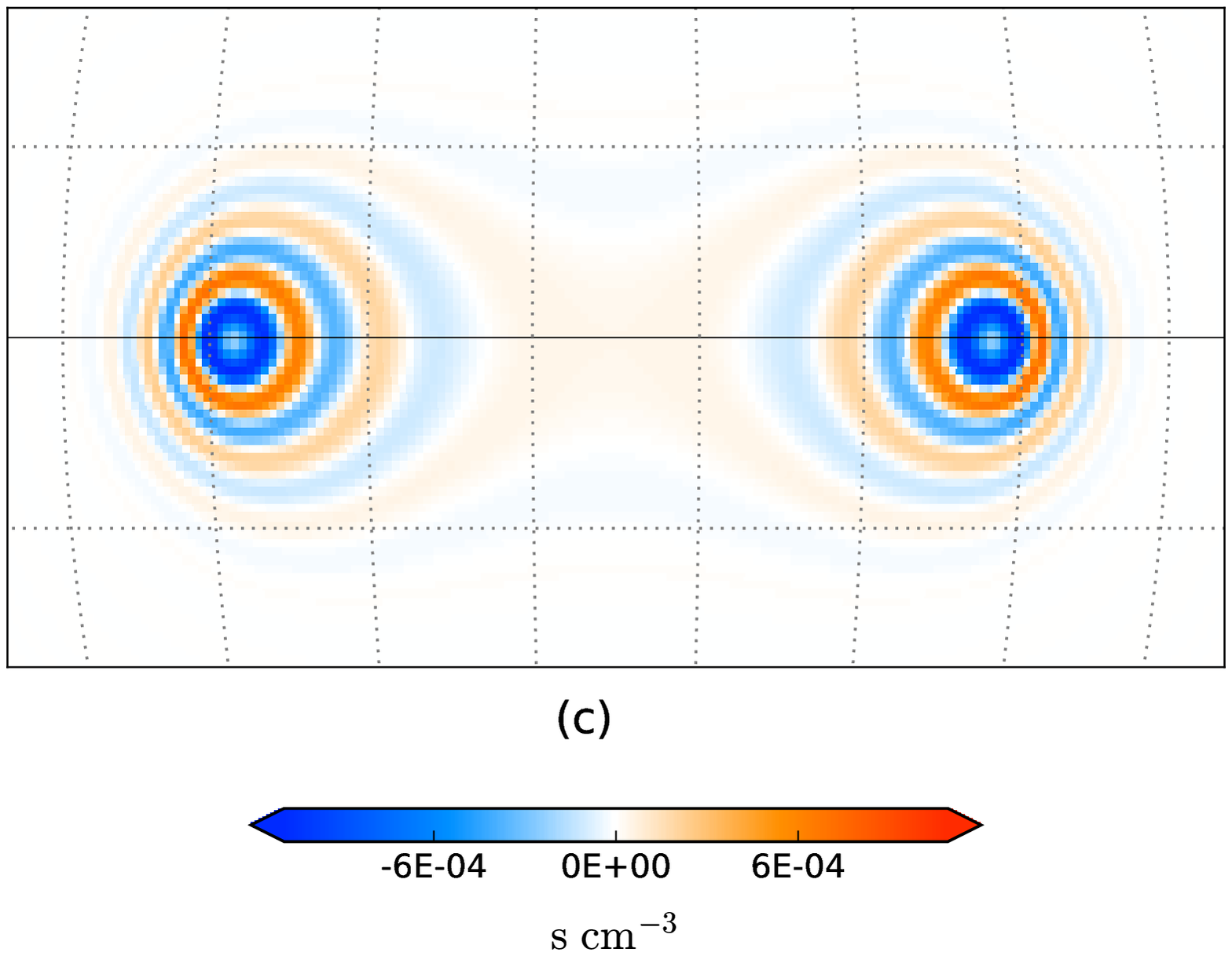}

\includegraphics[scale=0.45]{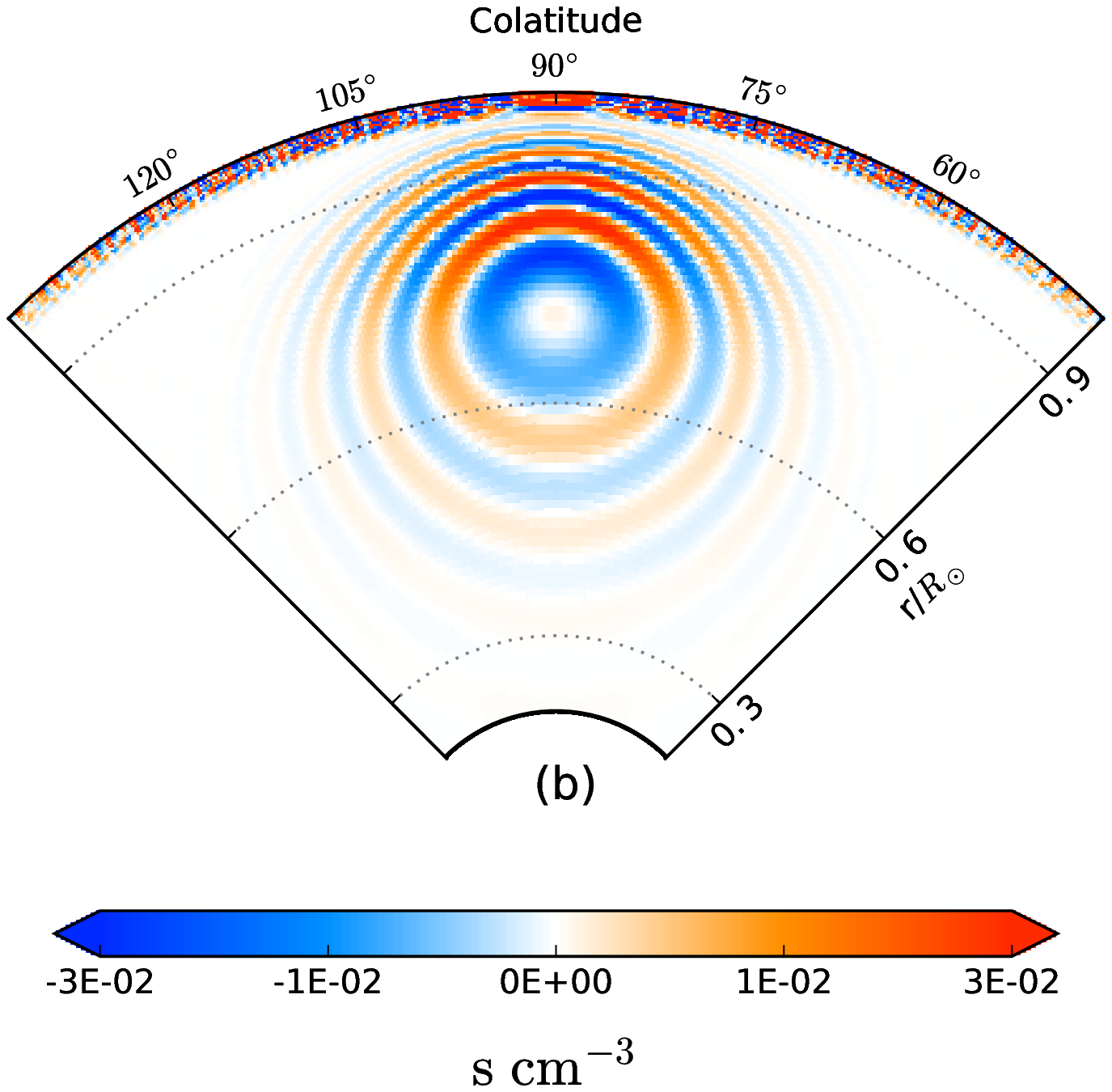}

\caption{\label{fig:sound_speed_kernel}Plot of $cK_{c}$ with source and receiver
$45^{\circ}$ apart, both placed on the equator. Sound-speed kernel
$K_{c}$ has been multiplied by sound-speed $c$ in order to magnify
the details of the sound-speed kernel in depth. Panel (a): Cut through
a plane containing both source and receiver and the center of the
Sun. The black dashed line connecting source and receiver is the ray
path evaluated for a frequency of 3.2 mHz. Panel (b): Cut through
a plane perpendicular to the ray path midway between source and receiver.
Panel (c): Slice of the kernel at $r=0.947\,R_{\odot}$ . We use the
Mercator projection for this plot. Longitudes and latitudes are represented
by dotted lines and equator by a black solid line. In all of the plots,
values of the kernels have been saturated to highlight details.}
\end{figure*}

We assume that the wave propagates through a background that has a
sound speed given by 
\begin{equation}
c\left(\mathbf{r}\right)=c_{0}\left(r\right)+\delta c\left(\mathbf{r}\right),
\end{equation}
where $\delta c\left(\mathbf{r}\right)$ is a small three-dimensional
perturbation to the spherically symmetric sound speed $c_{0}\left(r\right)$
in Model S. The corresponding change $\delta\mathcal{L}$ in the wave
operator $\mathcal{L}$ takes the form 
\begin{equation}
\delta\mathcal{L}\mathbf{G}_{r}=-2\bm{\nabla}(\rho_{0}c\delta c\bm{\nabla}\cdot\mathbf{G}_{r}),\label{eq:sound_speed_pert}
\end{equation}
where $\mathbf{G}_{r}=(G_{rr},G_{\theta r},G_{\phi r})$. Substituting
Equation (\ref{eq:sound_speed_pert}) in Equation (\ref{eq:gen_expn}),
we obtain the expression for the sound-speed kernel
\begin{equation}
K_{c}(\mathbf{r})=\int_{-\infty}^{^{\infty}}d\omega\,2\rho_{0}c\bm{\nabla}\cdot\mathbf{G}_{r}(\mathbf{r},\mathbf{r}_{r},\omega)\bm{\nabla}\cdot\mathbf{G}_{r}(\mathbf{r},\mathbf{r}_{s},\omega)h^{*}(\omega)F(\mathbf{r}_{s,}\omega).\label{eq:sound_speed_kernel}
\end{equation}
We have used the reciprocity relation derived from the adjoint nature
of the operator \citep{Shravan_2011} 
\begin{equation}
G_{ij}(\mathbf{r}_{1},\mathbf{r}_{2},\omega)=G_{ji}(\mathbf{r}_{2},\mathbf{r}_{1},\omega),\label{eq:reciprocity}
\end{equation}
to arrive at the Equation (\ref{eq:sound_speed_kernel}). The expression
for kernel $K_{c}$ is symmetric on the interchange of the source
and receiver location, $\mathbf{r}_{s}$ and $\mathbf{r}_{r}$ and
this symmetry can be seen in Fig. \ref{fig:sound_speed_kernel}. The
value of the kernel is small near the ray path --- as seen in ``banana-doughnut''
kernels in geophysics literature \citep{1999GeoJI.137..805M} ---
and peaks near the source and receiver locations. Fresnel zones surrounding
the ray path oscillate between positive and negative values. 

\subsection{Validation of sound-speed kernel \label{sound_speed}}

\begin{figure}[H]
\centering{}\includegraphics[scale=0.5]{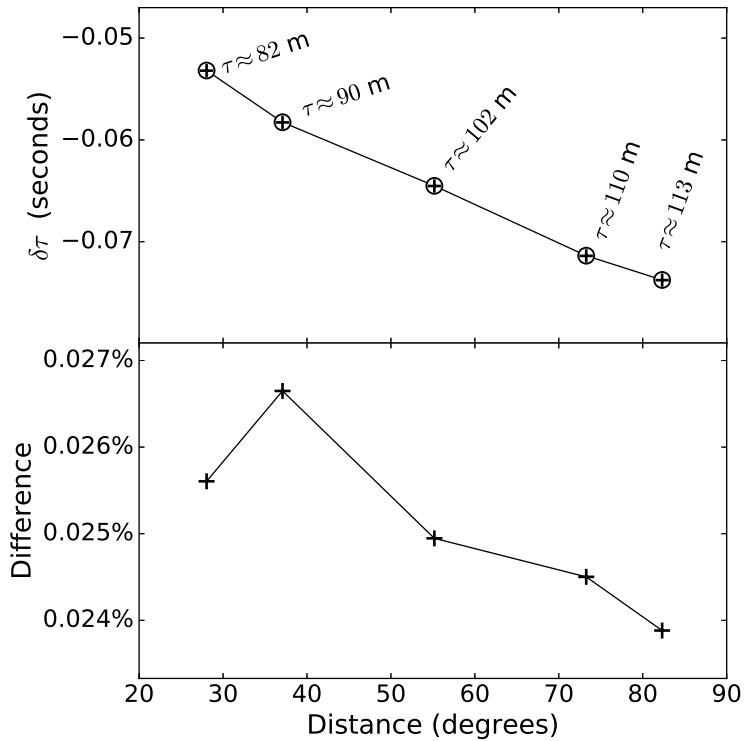}\caption{\label{fig:sound_speed_test}Upper panel: Comparison of travel-time
differences computed using the sound-speed kernel as $\delta\tau=\int d\mathbf{r}\,K_{c}\left(\mathbf{r}\right)\delta c\left(\mathbf{r}\right)$
(`+' symbols) and that computed from Equation (\ref{eq:travel_time})
when the sound-speed of model S is perturbed by $10^{-3}\%$. Approximate
travel time of the wave, $\tau$ (in minutes) is mentioned in the
plot alongside the points. The percentage difference between those
two values is plotted in the lower panel. }
\end{figure}

\begin{figure}
\begin{centering}
\includegraphics[scale=0.6]{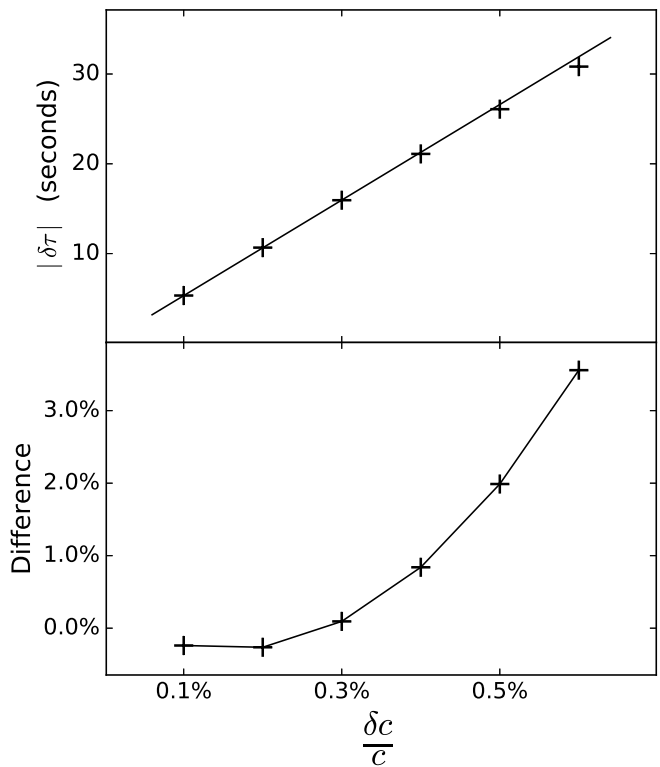}
\par\end{centering}

\caption{\label{fig:vary_sound_speed}Change in travel time due to perturbation
in sound-speed. Source and receiver are separated by an angular distance
of $27^{\circ}$. Upper panel: comparison of change in travel time
estimated from Equation (\ref{eq:travel_time}) (`+' symbols) and
that computed using sound-speed kernel (solid line) by varying the
perturbation in the sound speed of model S. Lower panel: relative
difference between the values computed through the two techniques
mentioned above. The mismatch increases with the magnitude of perturbation
because the first-order Born approximation loses validity.}

\end{figure}

In order to validate the sound-speed kernel, we consider the simple
scenario where the perturbation in sound speed $\delta c$ is only
function of the radius $r$, and the background remains spherically
symmetric. In that case, we can solve for Green's function numerically
in a manner similar to that described in Section \ref{green}, the
only change being $c_{0}\rightarrow c_{0}+\delta c$. After obtaining
the Green's function for the perturbed model, we can compute the $\xi_{r}(\mathbf{r}_{r},\mathbf{r}_{s},t)$
from Equation (\ref{eq:source_freq}). We also obtain the wave displacement
$\xi_{r}^{0}(\mathbf{r}_{r},\mathbf{r}_{s},t)$ for Model S through
a similar computation. Once we have both $\xi_{r}(\mathbf{r}_{r},\mathbf{r}_{s},t)$
and $\xi_{r}^{0}(\mathbf{r}_{r},\mathbf{r}_{s},t)$, we estimate the
change in travel time from Equation (\ref{eq:travel_time}) and compare
it with that obtained from sound-speed kernel (\ref{eq:kernel_travel_time}).
In Fig.~ \ref{fig:sound_speed_test}, we plot the results for several
different distances between source and receiver for a particular case
in which the sound speed of the model is perturbed by $10^{-3}\%$.
We find the two estimates of $\delta\tau$ to be in good agreement,
demonstrating that the sensitivity kernel has been computed accurately.
We also compare the accuracy of the sound-speed kernel by varying
the perturbation in sound speed in Fig. (\ref{fig:vary_sound_speed}). 

\subsection{Sensitivity kernel for flow\label{sub:sec-flow-kernel}}

\begin{figure*}
\includegraphics[scale=0.45]{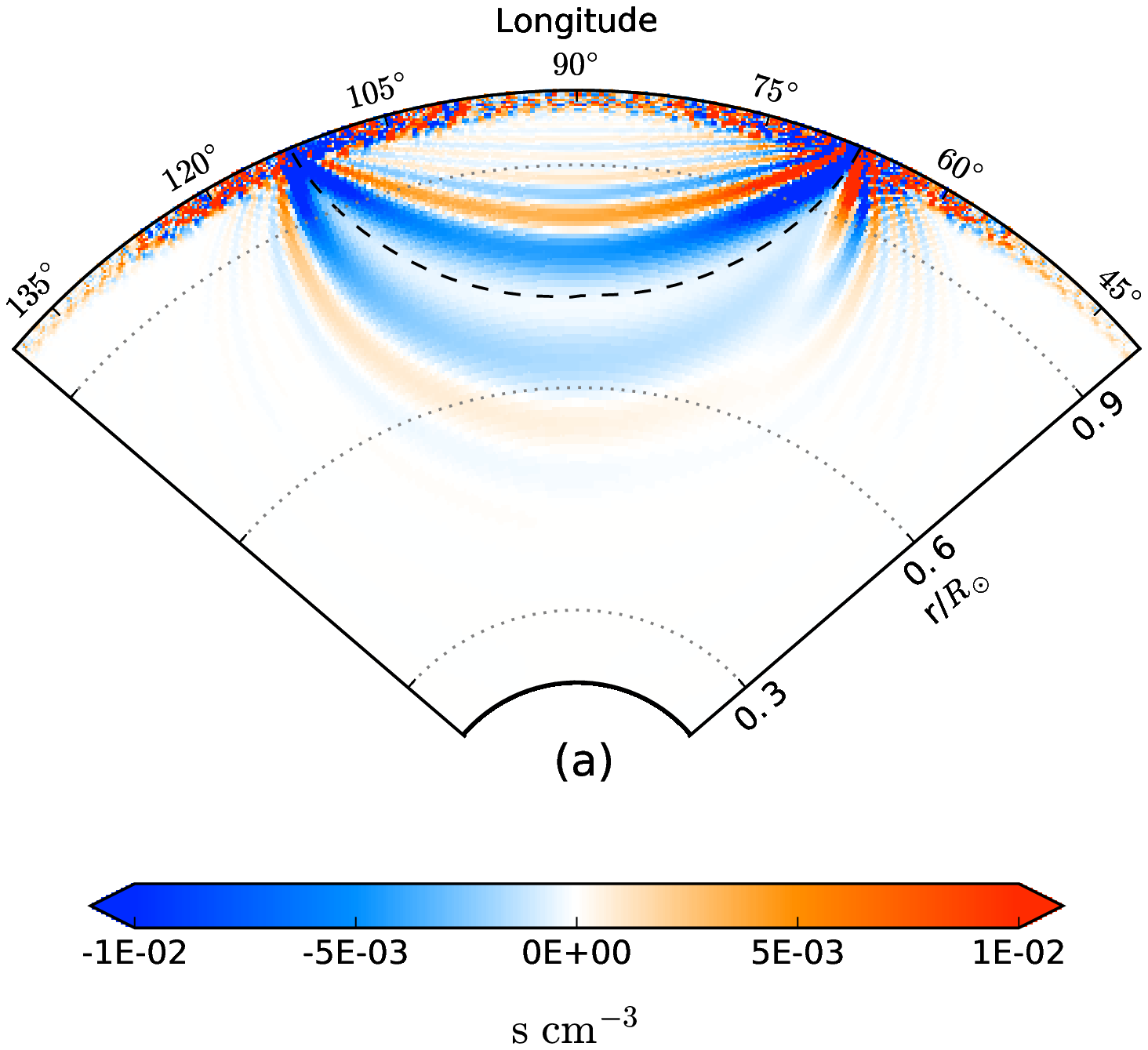}\includegraphics[scale=0.45]{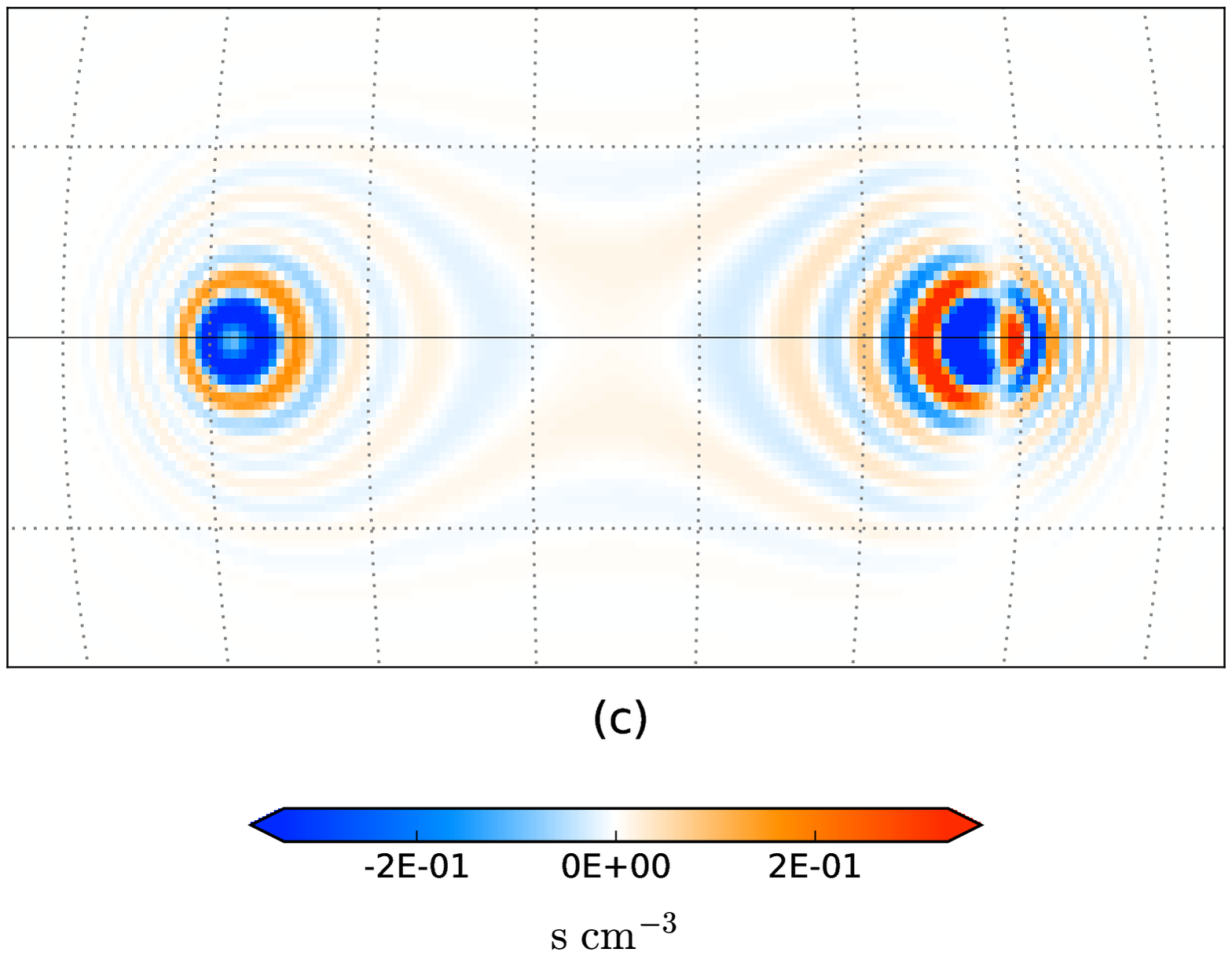}

\begin{raggedright}
\includegraphics[scale=0.45]{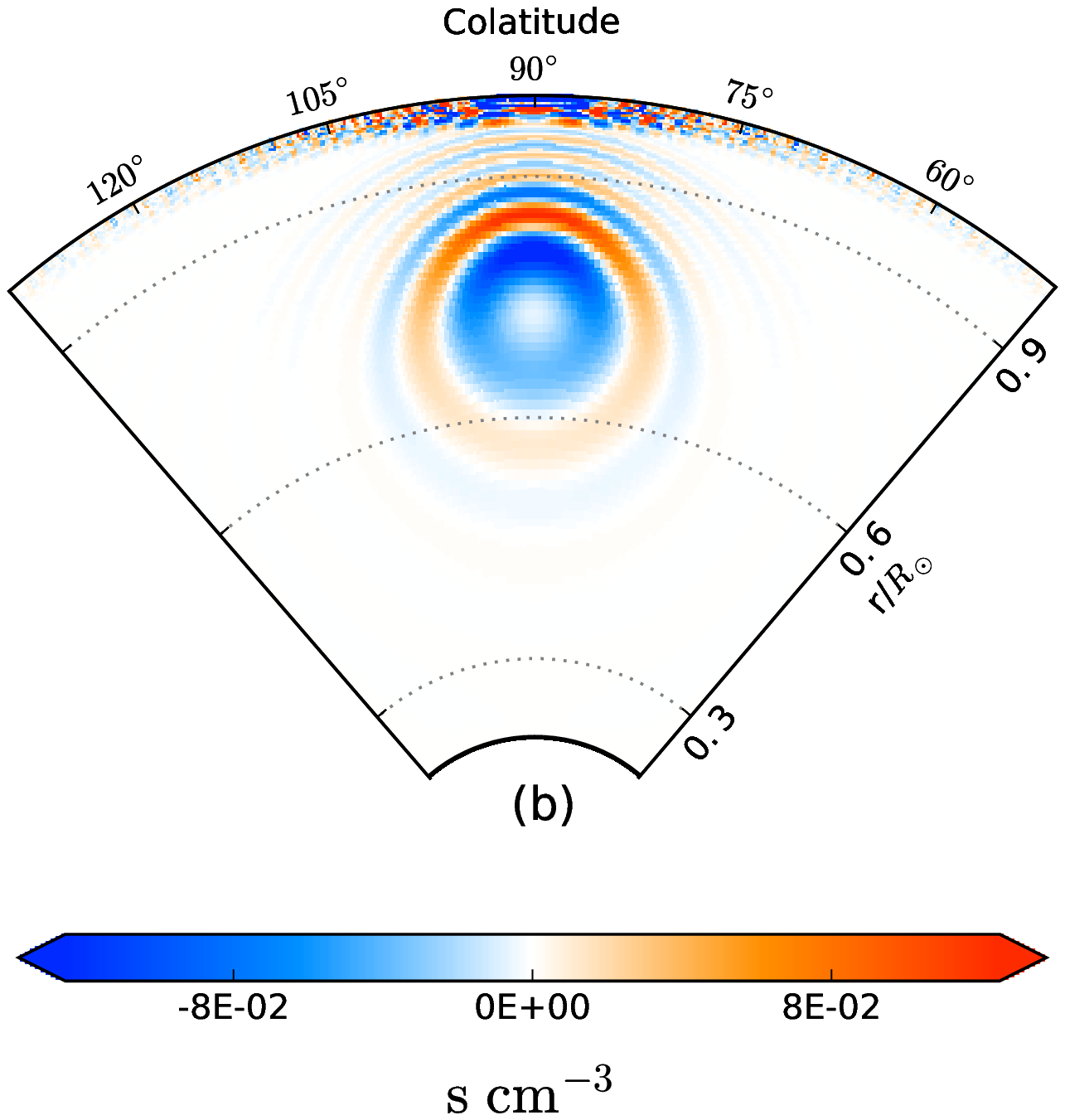}
\par\end{raggedright}

\caption{\label{fig:flow_kernel_phi}Plot of $cK_{v_{\phi}}$ where $K_{v_{\phi}}$
is the $\phi$ component of the sensitivity kernel for flow. Panel
(a): Cut through the plane containing source and receiver. Source
and receiver are separated by an angular distance of $45^{\circ}$.
The ray path, connecting source and receiver is shown by black dashed
line. Panel (b): Cut through the plane perpendicular to the ray path
at an equal distance from source and receiver. Panel (c): Slice of
the kernel at $r=0.947\,R_{\odot}$ . Mercator projection has been
considered for this particular plot. Longitudes and latitudes are
represented by dotted lines and equator by a black solid line. In
all panels, values of the kernels have been saturated to highlight
details.}
\end{figure*}

In presence of a temporally stationary flow with a velocity field
$\mathbf{v}\left(\mathbf{r}\right)$, there will be an advection term
in the wave equation given by 
\begin{equation}
\delta\mathcal{L}\bm{\xi}(\mathbf{r},\omega)=-2i\omega\mathbf{v}\cdot\bm{\nabla}\bm{\xi}(\mathbf{r},\omega).\label{eq:flow_pert}
\end{equation}
If the velocity field $\mathbf{v}$ is small compared to the sound-speed
$c$, the change in travel time $\delta\tau$ is linearly related
to $\mathbf{v}$,
\begin{equation}
\delta\tau=\int_{\odot}d\mathbf{r}\,\mathbf{K}_{\mathbf{v}}\left(\mathbf{r}\right)\cdot\mathbf{v\left(\mathbf{r}\right)},\label{eq:flow_time}
\end{equation}
where $\mathbf{K}_{\mathbf{v}}$ is the sensitivity kernel for velocity.
The expression for $\mathbf{K}_{\mathbf{v}}$ --- in the first-order
Born approximation --- is 
\begin{equation}
\mathbf{K}_{\mathbf{v}}=\int_{\odot}d\mathbf{\omega}\,2i\omega\rho_{0}G_{jr}(\mathbf{r},\mathbf{r}_{r})\bm{\nabla}G_{jr}(\mathbf{r},\mathbf{r}_{s})h^{*}(\omega)F(\mathbf{r}_{s},\omega),\label{eq:flow_kernel_expn}
\end{equation}
where index $j$ is summed over. We compute the $\theta$ and $\phi$
component of the velocity kernel $\mathbf{K}_{\mathbf{v}}$. The basic
features of the flow kernel are same as the sound-speed kernel. The
expression of kernel $K_{\mathbf{v}}$ is not symmetric in the source
and receiver locations, $\mathbf{r}_{s}$ and $\mathbf{r}_{r}$ and
this asymmetry is reflected in Fig. (\ref{fig:flow_kernel_phi}).
The flow kernel also has a small value along the ray path. 

\begin{figure}
\includegraphics[scale=0.45]{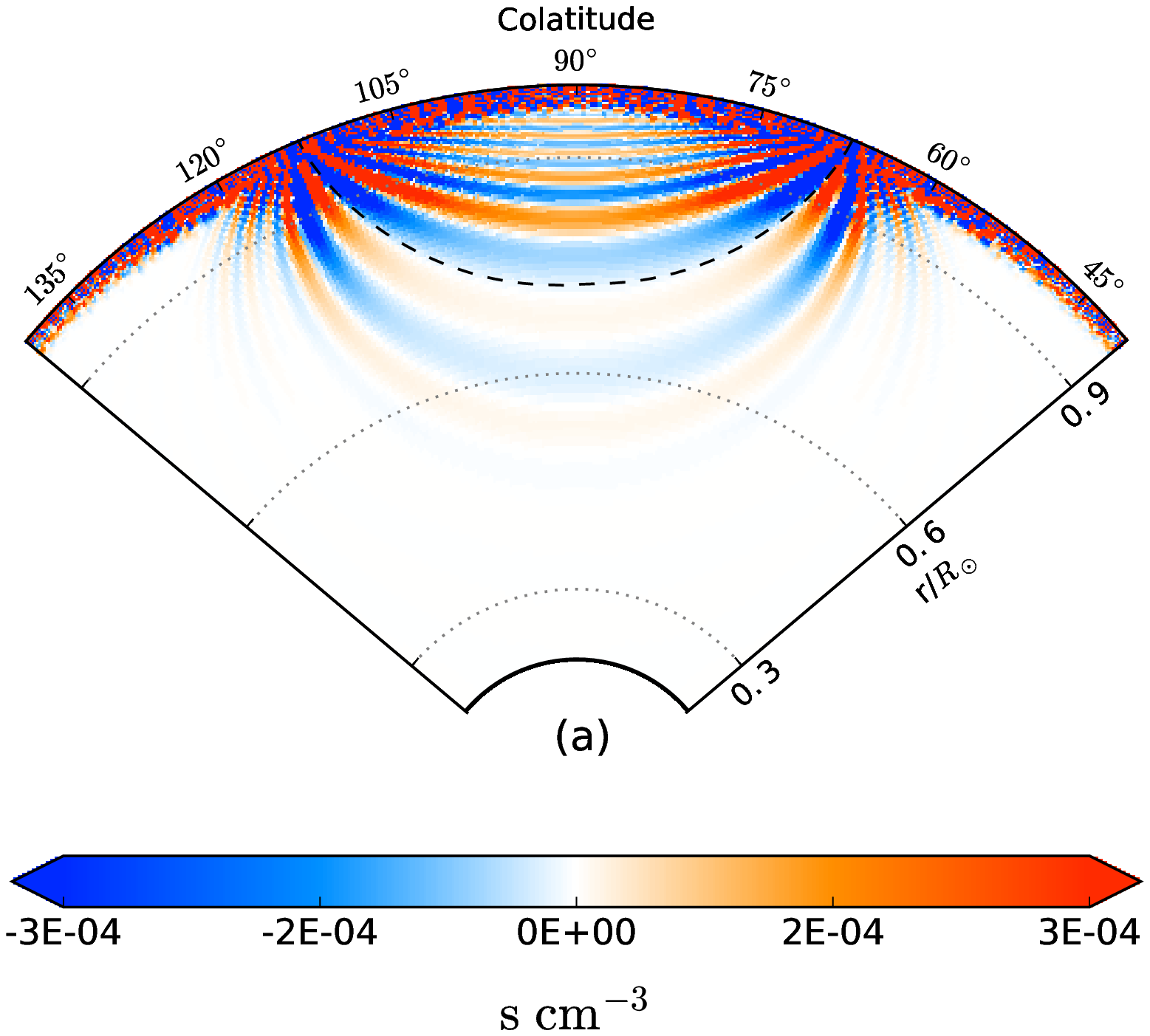}\includegraphics[scale=0.45]{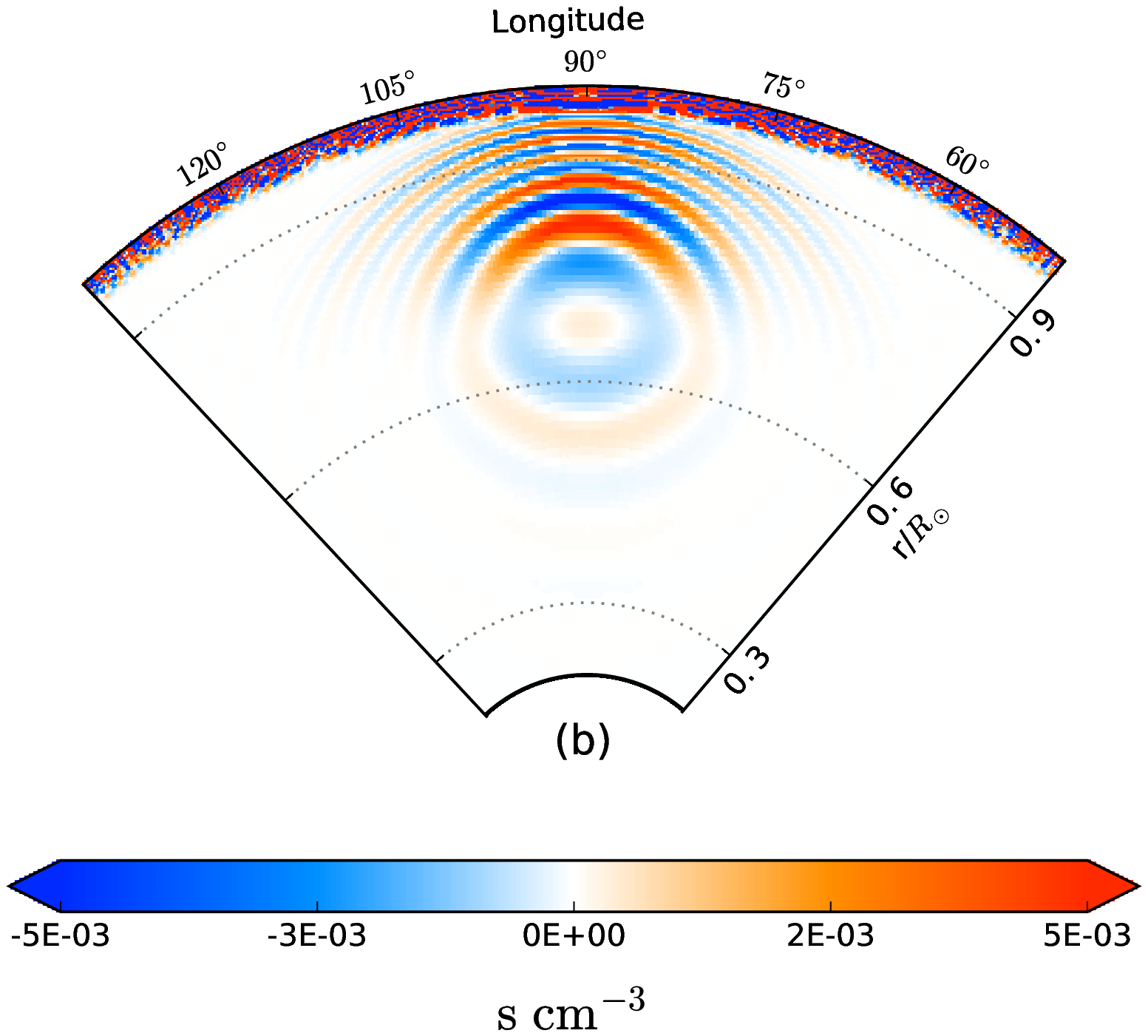}

\begin{centering}
\includegraphics[scale=0.3]{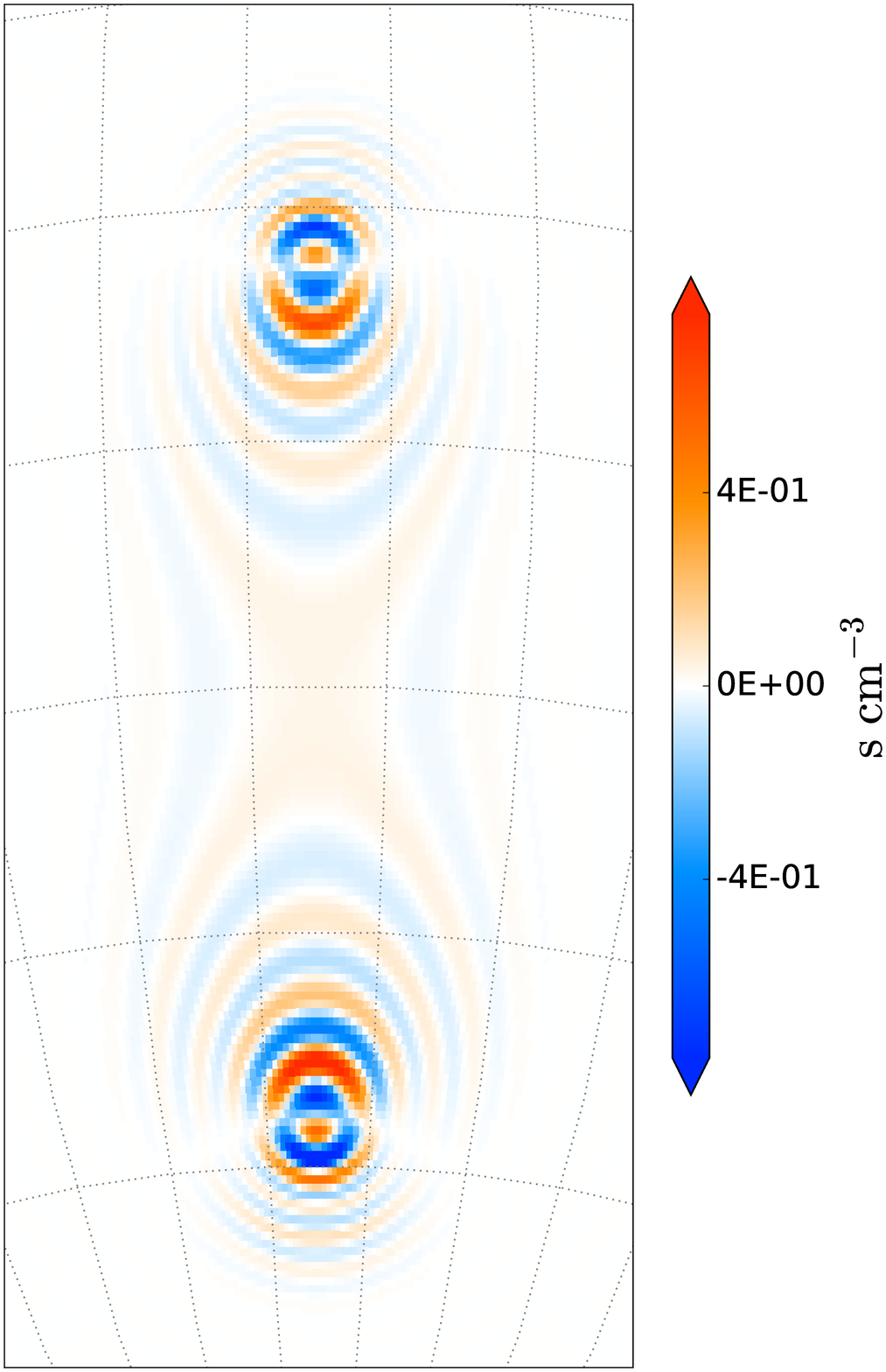}
\par\end{centering}

\caption{\label{fig:stream}Plot of $\rho K_{\chi}$. Source and receiver are
located on a meridian at an angular distance of $45^{\circ}$. Panel
(a): Cut through a plane containing source and receiver. The ray path
connecting source and receiver is shown by the black dashed line.
Panel (b): Cut through the plane perpendicular to the ray path, midway
between source and receiver. In all panels, values of the kernels
have been saturated to highlight details. Panel (c): Slice of the
kernel at $r=0.947\,R_{\odot}$ . Here also we have used Mercator
projection. Latitudes and longitudes are represented by dotted lines. }
\end{figure}

Realistic inversions for flows in the Sun should ensure mass conservation.
In temporally stationary backgrounds the condition for mass conservation
can be expressed as $\bm{\nabla}\cdot\text{\ensuremath{\left(\rho\mathbf{v}\right)}}=0$.
The constraint can be enforced automatically if we derive the velocity
field from a stream function $\bm{\chi}$. As we are interested in
meridional circulation, we follow the approach of \citet{2015ApJ...813..114R}
and consider an azimuthal stream function $\bm{\chi}=\chi\left(r,\theta\right)\hat{\bm{\phi}}$.
The corresponding velocity field is 
\begin{equation}
\rho\mathbf{v}=\bm{\nabla}\times\left(\chi(r,\theta)\hat{\bm{\phi}}\right).\label{eq:streamfn}
\end{equation}
We assume that $\chi=0$ at the solar surface. Substituting Equation
(\ref{eq:streamfn}) in Equation (\ref{eq:flow_time}), we obtain
\begin{eqnarray}
\delta\tau & = & \int_{\odot}d\mathbf{r}\,\chi\hat{\bm{\phi}}\cdot\bm{\nabla}\times\left(\frac{1}{\rho}\mathbf{K}_{\mathbf{v}}\right),\nonumber \\
 & = & \int_{\odot}d\mathbf{r}\,\chi K_{\chi}
\end{eqnarray}
where $K_{\chi}=\hat{\bm{\phi}}\cdot\bm{\nabla}\times\left(\frac{1}{\rho}\mathbf{K}_{\mathbf{v}}\right)$
is the sensitivity kernel for the stream function. We have computed
$K_{\chi}$ and it is shown is Fig. (\ref{fig:stream}). The values
of this kernel increases rapidly close to the surface, therefore we
multiply it with density before plotting to highlight the functional
variation with depth. The kernel, $K_{\chi}$ is shown in Fig (\ref{fig:stream}).
The grainy pattern near the surface is reminiscent of those observed
by \citet{2016ApJ...824...49B} and \citet{2016arXiv161101666G}.
It appears due to the finite cutoff in $\ell_{\text{max}}$ chosen
to compute the Green's function. Increasing $\ell_{\text{max}}$ appears
to further localize the pattern to shallower layers.

\subsection{$ $Validation of flow kernel}

\begin{figure}[H]
\begin{centering}
\includegraphics[scale=0.6]{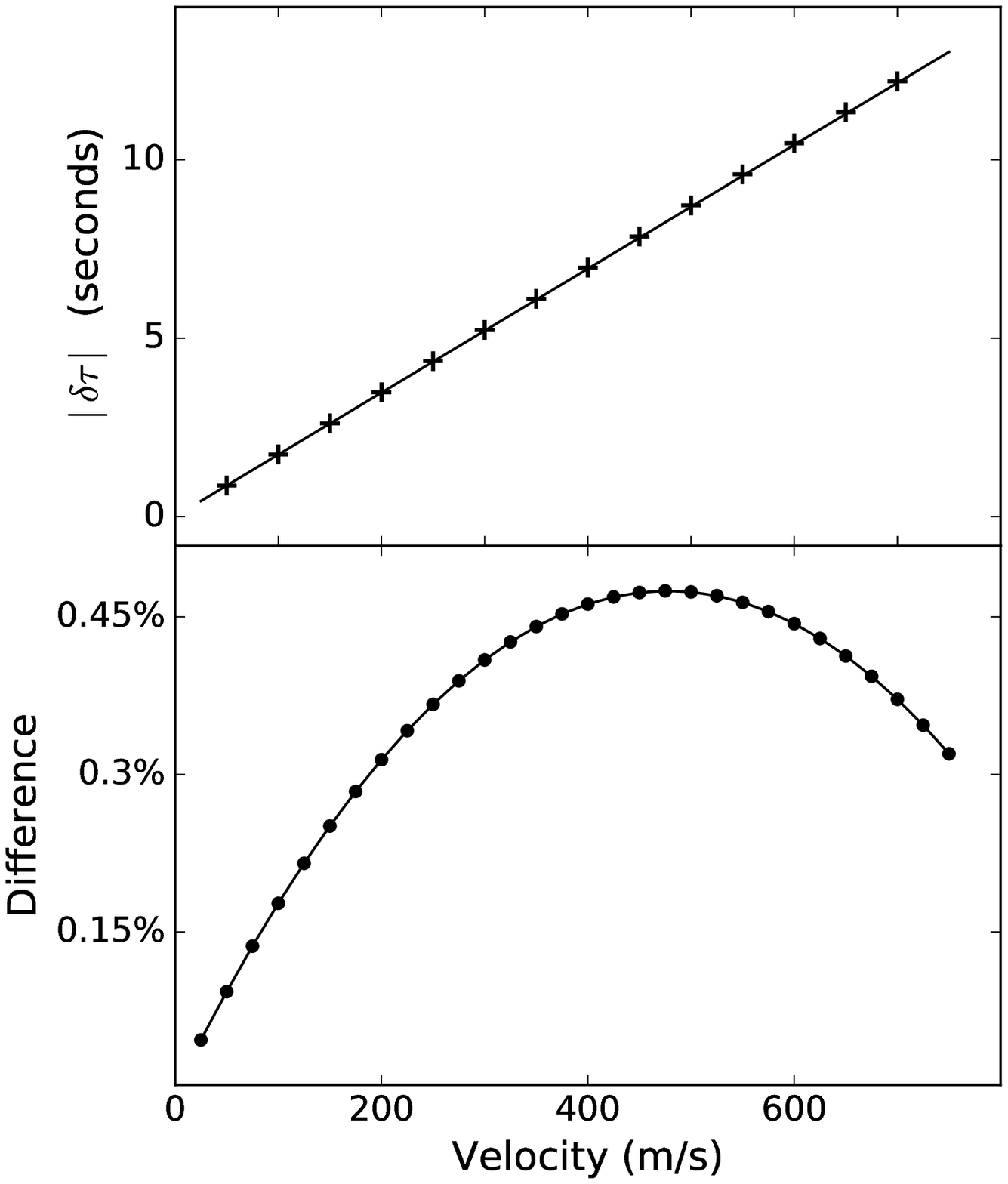}
\par\end{centering}

\caption{\label{fig:flow_kernel_test} Change in travel time due to uniform
rotation of the Sun. The source and receiver, $50.6^{\circ}$ apart
are both placed on the equator . Upper panel: change in travel time
obtained from Equation (\ref{eq:travel_time}) (`+' symbols) and from
flow-kernel (Equation (\ref{eq:flow_time}), solid line). Lower panel:
relative difference between change in travel time obtained from Equation
(\ref{eq:travel_time}) and flow kernel. }
\end{figure}

\begin{figure}
\centering{}\includegraphics[scale=0.5]{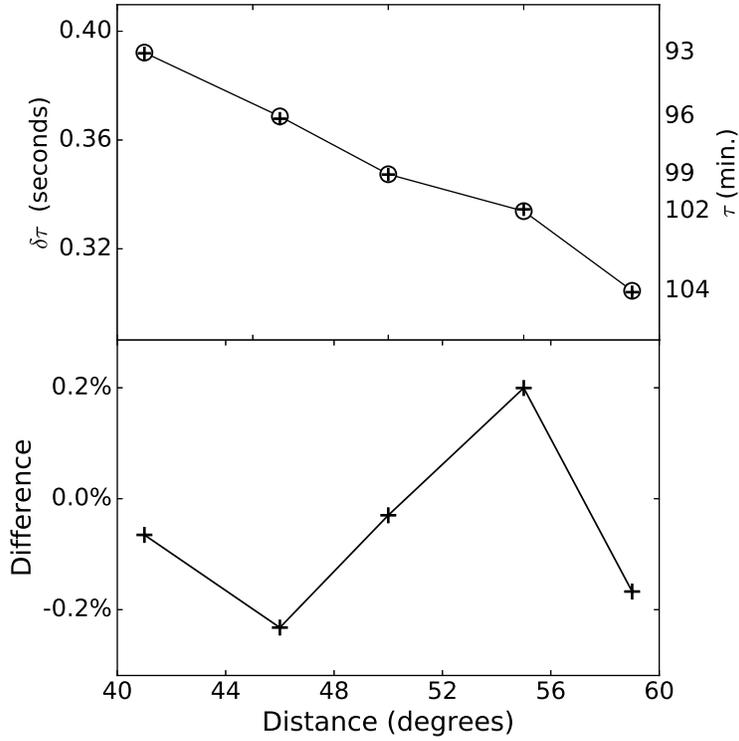}\caption{\label{fig:kernel_distance_flow}Upper panel: comparison of change
in travel time $\delta\tau$ estimated from Equation (\ref{eq:flow_time})
( `+' symbols) with that estimated from Equation (\ref{eq:travel_time})
(solid line with circles) for different source-receiver distances
when the surface flow speed is $20$ m/s. Corresponding travel times
$\tau$ are indicated on the right. Lower panel: relative difference
between the values computed through the two techniques mentioned above. }
\end{figure}

To test the accuracy of the kernel, we consider a flow field equivalent
to a solid-body rotation, thereby retaining spherical symmetry. In
this case the velocity field will have the form
\begin{equation}
\mathbf{v}(\mathbf{r})=\Omega r\sin\theta\hat{\bm{\phi}},\label{eq:velocity_phi}
\end{equation}
where $\Omega$ is the angular velocity of the rotation. The perturbed
wave field $\bm{\xi}$ will be related to unperturbed wave field $\bm{\xi}_{0}$
through a change in reference frame
\begin{equation}
\xi_{r}(\Delta,t)=\xi_{r,0}(\Delta-\Omega t,t),\label{eq:rotation}
\end{equation}
where $\Delta$ is the angular distance between source and receiver.
We place both the source and receiver on the equator. Using Equation
(\ref{eq:rotation}), we compute $\delta\tau$ from Equation (\ref{eq:travel_time})
and compare it with that obtained from Equation (\ref{eq:flow_time}).
We plot the dependence $\delta\tau$ on the strength of flow velocity
in Fig. \ref{fig:flow_kernel_test}. In Fig. \ref{fig:kernel_distance_flow},
we show the dependence of $\delta\tau$ with source-receiver distance
when the flow speed is $20\,\text{m/s}$ at the surface, and compare
the difference between the values computed through the two techniques
mentioned above. 

\section{Summary and discussion\label{conclusion}}

In this paper, we have developed a technique to compute seismic sensitivity
kernels in spherical geometry using the first-order Born approximation.
Computation of spherical sensitivity kernels are typically expensive.
We have shown that assuming a spherically symmetric background, Green's
function decouples in frequency and harmonic degree and therefore
computation of each frequency and harmonic degree can be done efficiently
in parallel on a computer cluster. It takes around $16$ seconds to
compute the displacement vector and pressure perturbation of equation
(\ref{eq:matrix_eqn}) for each $(\ell,\,\omega)$ pair on a single
processor. For the parameters chosen in this work, the entire Green's
function takes around six hours to compute when evaluated in parallel
using $300$ processors on a computer cluster. It takes a further
hour to compute the sensitivity kernel from the Green's function.

We have studied in this work how weak flow has to be in order for
linear relationship between travel-time delay and flow to hold. We
have found that travel times can be obtained within $0.47\%$ accuracy
using the flow kernel computed through our approach for uniform flows
up to $750$ m/s. Since the observed velocity of meridional circulation
on the solar surface is around $20$ m/s, we expect that linearity
might be an appropriate assumption for the study of meridional circulation. 

We have considered a single deterministic source in our work.\textbf{
}In the case of uniformly distributed sources and for certain types
of wave damping, it can be shown that \citep{snieder2004extracting,snieder2007extracting}
the positive and negative branches of the cross-correlation measurement
may be interpreted as waves originating from one measurement pixel
to the other and vice versa. This equivalence between cross correlations
and Green's function, while possibly not very accurate in the Sun
owing to line-of-sight projection and a complicated damping mechanism
(among other effects), represents a useful starting point. Indeed,
travel-time inversions of meridional circulation are typically performed
using kernels computed in the ray approximation (e.g. \citet{1997Natur.390...52G,2013ApJ...774L..29Z,2015ApJ...813..114R}).
Ray theory assumes that the wave frequency is infinite, relies on
a single-source picture and does not take into account line-of-sight
projection. In contrast, the Born approximation can account for line-of-sight
projection and because it is a finite-frequency model, is more accurate
than ray theory. Therefore kernels based on the Born approximation,
computed in the single-source picture, though not the best, are still
better to use for inversions than kernels computed using ray theory.
A more complete theory would aim to model the cross-correlation measurement
and take into account line-of-sight projection effects, which will
be a part of our future work. 

KM, JB \& SMH acknowledge the financial support provided by the Department
of Atomic Energy, India. SMH also acknowledges support from Ramanujan
fellowship SB/S2/RJN-73/2013, the Max-Planck partner group program
and thanks the Center for Space Science, New York University at Abu
Dhabi. 

\bibliographystyle{apj}
\bibliography{krishnendu}

\end{document}